\documentclass{llncs}
\usepackage{booktabs} 
\usepackage{amsmath}
\usepackage{booktabs}
\usepackage{amssymb}
\usepackage{multirow}
\usepackage{pgf,pgfplots}
\usepackage{tikz,pgfplotstable}
\pgfplotsset{compat=1.7}
\usepackage{xcolor}
\usepackage{pgfplots}
\usepackage{tikz}
\usepackage{graphicx}
\usepackage{subfigure}
\usepackage{float}
\usepackage{epstopdf}
\usepackage{amsmath}
\usepackage{comment}
\usepackage{algorithm}
\usepackage{algpseudocode}
\usepackage{pifont}
\usepackage{blkarray}

\usepackage{enumitem}
\usepackage{flushend}
\usepackage{caption}
\usepackage[english]{babel}
\usepackage{wrapfig,lipsum}

\begin{document}
\title{Peacock: Probe-Based Scheduling of Jobs by Rotating Between Elastic Queues}
\author{Mansour Khelghatdoust \and Vincent Gramoli}
\institute{The University of Sydney, Australia}
\maketitle              
\begin{abstract}
In this paper, we propose Peacock, a new distributed probe-based scheduler which handles heterogeneous workloads in data analytics frameworks with low latency. Peacock mitigates the \emph{Head-of-Line blocking} problem, i.e., shorter tasks are enqueued behind the longer tasks, better than the state-of-the-art. To this end, we introduce a novel probe rotation technique. Workers form a ring overlay network and rotate probes using elastic queues. It is augmented by a novel probe reordering algorithm executed in workers. We evaluate the performance of Peacock against two state-of-the-art probe-based solutions through both trace-driven simulation and distributed experiment in Spark under various loads and cluster sizes. Our large-scale performance results indicate that Peacock outperforms the state-of-the-art in all cluster sizes and loads. Our distributed experiments confirm our simulation results. 
\keywords{Scheduling, Distributed System, Load Balancing, Big Data}

\end{abstract}
\section{Introduction}
Data analytics frameworks increase the level of parallelism by breaking jobs into a large number of short tasks operating on different partitions of data to achieve low latency. Centralized techniques schedule jobs optimally by having near-perfect visibility of workers. However, with the growth of cluster sizes and workloads, scheduling time becomes too long to reach this optimality. To solve this problem, probe-based distributed techniques have been proposed~\cite{ACM:Ousterhout,Eagle,Hawk} to reduce the scheduling time by tolerating a suboptimal result. These solutions typically sample two workers per probe and place the probe into the queue of the least loaded worker. Additionally, they are augmented with amelioration techniques such as re-sampling, work stealing or queue reordering to likely improve the initial placement of probes. However, the existing algorithms are not able to improve scheduling decisions continuously and deterministically to mitigate the \emph{Head-of-Line blocking}, i.e., placing shorter tasks behind longer tasks in queues, efficiently. Moreover, the overall completion time of a job is equal to the finish time of its last task. Due to the distributed and stateless nature of probe-based schedulers, the existing solutions are not able to reduce the variance of tasks completion time of each job that are scheduled on various workers to reduce job completion time.     

We propose Peacock, a fully distributed probe-based scheduler, which replaces the probe sampling and the unbounded or fixed-length worker-end queues with a deterministic probe rotation and elastic queues. This leads to better scheduling decisions while preserving fast scheduling of jobs. This probe rotation approach finds an underloaded worker better than probe sampling because probes traverse a higher number of workers.
Workers are organized into a ring and send probes to their neighbors at fixed intervals. A probe rotation lets a loaded worker delegates the execution of a probe to its successor on the ring. \emph{Elastic queues} regulate the motion of probes between workers and lets a worker dynamically adjust its queue size to balance load between workers. By decreasing the queue size, workers are forced to move some of their probes and increase the queue size to avoid unnecessary motion of probes. More interestingly, a probe in its journey, from when it is submitted to the scheduler until it runs on any arbitrary worker, moves between workers, stays in some worker and then continue rotating until eventually executing on a worker. Furthermore, Peacock is augmented with a probes reordering to handle the \emph{Head-of-Line blocking} more effectively. The probes of one job are annotated with an identical threshold time equals to the cluster average load at the time of scheduling. This threshold determines a soft maximum waiting time for probes that are scattered independently between workers to reduce the variance of job completion time. 

We evaluate Peacock through both simulation and distributed experiments. We use trace from Google~\cite{GoogleTraceWebsite}. We compare Peacock against Sparrow~\cite{ACM:Ousterhout} and Eagle~\cite{Eagle}, two state-of-the-art  probe-based schedulers. The results show Peacock outperforms Eagle and Sparrow in various cluster sizes and under different loads. We evaluate the sensitivity of Peacock to probe rotation and probe reordering. Section~\ref{Peacock} describes Peacock in details. Section~\ref{sec:Evaluation} explains the evaluation methodology. Section~\ref{sec:Simulation} describes simulation and implementation results. Section~\ref{sec:Relatedwork} discusses related work. Section~\ref{sec:Conclusion} concludes the paper.     
\section{The Peacock Scheduler}\label{Peacock}
\begin{wrapfigure}{r}{6.5cm}
\centering
\includegraphics[scale=0.4]{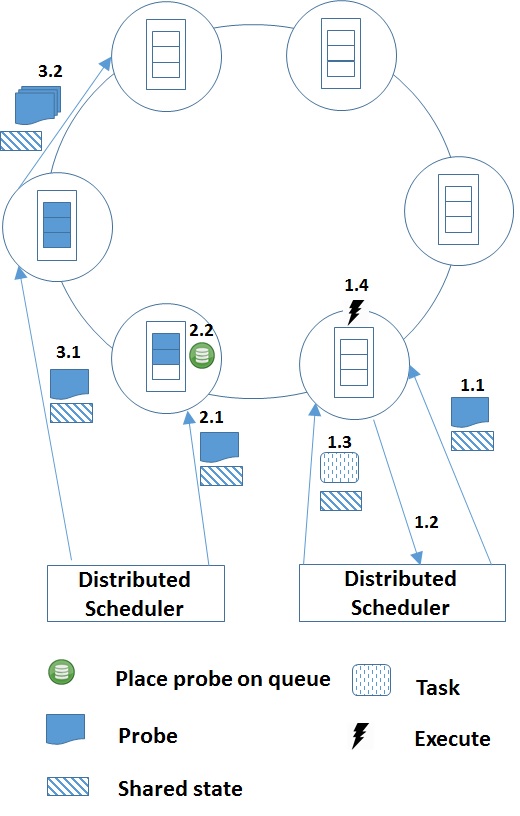}
\caption{Different scenarios workers handle probes.}
\label{fig:architecture}
\end{wrapfigure}

Peacock comprises a large number of workers and a few schedulers. Workers shape a ring overlay 
network in that each worker connects to its successor and additionally stores descriptors to a few successors for fault tolerance purpose. Each scheduler  connects to all workers. Schedulers manage the life cycle of each job without the need for expensive algorithms. Jobs are represented as a directed acyclic graph (DAG), with tasks as vertices and data flow between tasks as edges. This DAG is divided into stages and actually Peacock considers each stage as a job and hence a DAG consists of a number of dependent jobs. Similar to other approaches~\cite{Hawk,Eagle,Apollo,ACM:Rasley}, Peacock needs to know the estimated task runtime of incoming jobs which is measured by methods explained elsewhere~\cite{SCOPE,Apollo}. Jobs can be scheduled by any of the schedulers, however, all tasks of a job are scheduled by the same scheduler. When a scheduler has received a job, it submits probe messages to a set of random workers equals to the number of tasks. Each worker has a queue. According to the Figure~\ref{fig:architecture}, once a worker has received the probe, (a) if the worker is idle (1.1), it requests the corresponding task of the probe from the scheduler (1.2) and the scheduler sends back the corresponding task data (source code) (1.3) and then the worker executes the task (1.4), (b) if the worker is executing a task and its queue consists of a number of waiting probes like (2.1) and (3.1), the worker may enqueue the probe for the future execution or rotation (2.2), or (c) the worker may either rotate the incoming probe instantly or enqueue the probe and rotate other existing waiting probes (3.2).
\subsection{Probe Rotation}\label{sec:probeRotation}
There are three important design questions that should be answered: \begin{enumerate}[label=(\roman*)]
\item How should probes move between workers?
\item When should each worker rotate probes?  
\item Which probes should each worker choose to rotate? 
\end{enumerate}
\subsubsection{Ring Overlay Network.}
The challenging design decision is how probes move between workers. The easiest solution is that workers maintain a complete list of workers and send probe to a sampled worker. However, it undermines the scalability and burdens some workers while some others might remain mostly idle. The efficient approach should be symmetric, balance load between workers and maximize resource utilization. To this end, Peacock exploits a ring overlay network as depicted in Figure~\ref{fig:architecture}. We discuss whether exploiting a ring overlay network adversely impacts the scalability of Peacock. Peer-to-Peer overlay networks are extensively used to implement routing and lookup services~\cite{chord}. In this respect, applying a ring overlay network with 1 in-out degree (i.e., 1 for in-degree and 1 for out-degree) in which lookup time grows linearly with the increment of ring size ruins scalability. However, there is no routing or lookup service in Peacock. It only rotates probes through a ring and typically probes are able to execute on any arbitrary worker node. Schedulers submit probes to sampled workers  and probes are either rotated or stores at workers. Therefore, we can conclude that exploiting a ring overlay network does not undermine the scalability of the algorithm.
\subsubsection{The Significance of Elastic Queues.}\label{sec:elasticQueues}
Workers should decide when and which probes to rotate. Each worker utilizes one \emph{elastic} queue, i.e., the size is adjusted dynamically and hence is resilient. This elasticity is crucial for queues because it enables workers to rotate probes between themselves in order to distribute the probes uniformly. If queues are too short, the resources get under-utilized due to the existence of idle resources between allocations. If the queues are too long, then the load among workers gets imbalanced and job completion gets delayed. Determining a static queue size might lead to an excessive number of probe rotations when the cluster is heavily loaded and an inefficient reduction in the number of probe rotations when the cluster is lightly loaded. Peacock bounds queues using a pair ($<$\textit{size, average load}$>$) which is called \textit{shared state}. The size is calculated as the average number of current probes on cluster. The average load is calculated as the average estimation execution time of current probes on workers. This pair is adjusted dynamically to make queues resilient. \subsubsection{Shared State.}\label{shared state}
\emph{Shared state} is a pair of information that consists of the queue size and the average load of cluster ($<$\textit{queue size, average load}$>$) and is changing from time to time since the cluster has dynamic workload. Workers require to get the most recent \textit{shared state}. However, it is challenging to update  the \textit{shared state} of workers continuously in a decentralized manner. Peacock is designed in such a way that workers and schedulers are not strictly required to have an identical \textit{shared state} all the time and hence workers may have different values of \textit{shared state} at times. Now, we describe how the \textit{shared state} is calculated and through what ways workers can get the latest value of \textit{shared state}. Each scheduler calculates the \textit{shared state} continuously based on the messages it receives. These messages are when a scheduler receives a job arrival event, receives a task finish event \textbf{or} receives an update message from other schedulers. For example, suppose the current aggregation load of cluster is $<$1500, 25000$>$ (the number of probes, aggregation load) and a task finished event is received for a task with 20s estimated execution time. The scheduler updates the aggregation value to $<$1499, 24980$>$ and sends asynchronously the message $<$- , 1 , 20$>$ to the other schedulers. Upon receiving this message, the other schedulers update their aggregation value. Similarly, receiving a new job with 10 tasks and 15s estimated execution time changes the aggregation value to $<$1510, 25150$>$, with update message $<$+ , 10 , 150$>$ to the other schedulers. As an alternative solution, schedulers can manage \textit{shared state} through coordination services such as ZooKeeper. It eliminates direct communication between schedulers.    
Each scheduler calculates the value of \textit{shared state} through dividing aggregation value by the number of workers. Peacock does not impose extra messages to update the \textit{shared state} of workers. The latest \textit{shared state}  is piggybacked by messages that workers and schedulers exchange for scheduling purposes. Figure~\ref{fig:architecture} shows workers get \textit{shared state} through three ways.
\begin{enumerate}[label=(\roman*)]
\item When schedulers submit a probe message to workers 
\item When schedulers send task data as a response of getting task by worker
\item When workers rotate probes to their neighbors.  
\end{enumerate}
\subsubsection{Rotation Intervals.}\label{sec:rotation}
In ring,  workers rotate probes to their successor. Peacock rotates probes periodically in rounds. Once a probe has been chosen to be rotated, it is marked for rotation until the next round. In the next round, workers send all the marked probes in one message to their neighbors. Such design reduces the number of messages that workers exchange. Most jobs consist of a large number of probes and it is common that in each round more than one probe of the same job are marked by the same worker to rotate. Peacock leverages this observation to remove the redundant information of such subset of probes to reduce the size of messages. To reduce the number of messages, workers send rotation message to their neighbor only if either there is/are probe(s) marked for rotation or when the \textit{shared state} is updated from the last round. The interval between rounds is configurable from milliseconds to few seconds and it does not impact the job completion time since one probe is marked for rotation. This avoids having to wait 
in a long queue.
\subsection{Probes Reordering}\label{sec:probesReordering}
It is crucial to reduce the variance of probes queuing time of one job since job completion time is affected by the last executed task of the job. It is challenging since  probes of a job are distributed on different workers. However, the addition of the probes to queues in FIFO order (i.e., in the order in which they are arrived) does not decrease the queuing time variance in the presence of heterogeneous jobs and  workloads.  Probe reordering is a solution to this problem~\cite{Apollo,Eagle}. Reordering algorithms should ideally be starvation-free, i.e., no probe should starve due to existence of infinite sequence of probes with higher priority.
To this end, we propose a novel probe reordering algorithm. It performs collaboratively along with probe rotation algorithm to mitigate the \emph{Head-of-Line blocking}. Since probes rotate between workers, the algorithm cannot rely on  FIFO ordering of queues. Assume a scheduler submits probe $p_{1}$ to worker $n_{1}$ at time $t_{1}$  and probe $p_{2}$ to worker $n_{2}$ at time $t_{2}$. Then, $n_{1}$ rotates $p_{1}$ and reaches $n_{2}$ at time $t_{3}$. The problem is that $p_{1}$ is placed after $p_{2}$ in the queue of $n_{2}$ while it has been scheduled earlier. To overcome this problem, schedulers label job arrival time on probe messages so that workers place incoming probes into queues w.r.t the job arrival time. Then, schedulers attach task runtime estimation to probe messages. Once a worker has received a probe, it orders probes by giving priority to the probe with the shortest estimated runtime. While it reduces the \emph{Head-of-Line blocking}, it may ends in starvation of long  probes. To avoid this issue, schedulers attach a threshold value to all the probes of a job at arrival time. The value is the summation of the current time and the average execution time extracted from the current \textit{shared state}. For example, if one job arrives at $t1$ and the \textit{shared state} value is 10s threshold, the value is $t1$ + 10 for all probes of that job. This threshold acts as a soft upper-bound to reduce tail latency and hence to reduce job completion time. It avoids starvation since probes do not allow other probes to bypass them after exceeding the threshold time and hence they  eventually move to the head of queue and execute on worker.\\
We now present the algorithm. Workers receive a probe either because their predecessor rotates it along the ring or because the probe is submitted by a scheduler. Algorithm~\ref{algo:enqueue} depicts the procedure of enqueuing a probe and Table~\ref{table:notations} explains the associated notations. Peacock maintains a sorted queue of waiting probes. Once a new probe has arrived, it is treated as the lowest priority among all waiting probes (Line 2) and tries to improve its place in the queue by passing other probes. It starts comparing its arrival time with the lowest existing probe (Line 4). If the new probe has been scheduled later than the existing probe, bypassing is not allowed unless it reduces head-of-line blocking without leading to starvation of the comparing probe. Bypassing the new probe can mitigate the \emph{Head-of-Line blocking} if the execution time of the new probe is less than the existing probe. Such bypassing should not lead to the starvation of the passed probe which is checked through threshold. If the threshold of the existing probe has not exceeded in advance or will not exceed due to bypassing, then the new probe can bypass the existing probe. Otherwise, it is either simply enqueued or rotated to the neighbor worker on the ring (Lines 4-10). If the new probe has been scheduled earlier, it cannot bypass if the existing probe has less execution time. The new probe does not exceed the threshold if it does not bypass (Lines 11-16). Then, the new probe waits in the queue if it does not violate the starvation conditions, otherwise it is marked to be rotated in the next coming round (Lines 25-31). Once the process of enqueuing the probe has finished, Peacock checks the shared state of the worker and may rotate one or more probes if needed (Lines 21-23). 
\begin{table}[htp!]
\setlength{\tabcolsep}{1pt}
\caption{List of notations}
\label{table:notations}
\centering 
\begin{tabular}{    c p{5cm}     c p{5cm}     c p{5cm}     c p{5cm} } 
\toprule
  Symbol &    Description &    Symbol &      Description\\
\toprule
  $\phi$ & Queue Size & $\omega$ & Max threshold waiting probes   \\
  $\tau$ & Current time  & $\mu$ & Max threshold waiting time for $p$\\
  $\lambda$ & Job arrival time & $\theta$ & runtime estimation of probe $p$   \\
  $\alpha$ & Total runtime of waiting probes & $\beta$ & Arrival time probe $p$  \\
  $\gamma$ & Waiting time estimation probe $p$ & $\delta$ & Relict runtime of running task \\
\toprule
\end{tabular}
\end{table}
\section{Evaluation Methodology}\label{sec:Evaluation}
\textbf{Comparison.} We compare Peacock against Sparrow~\cite{ACM:Ousterhout} and Eagle~\cite{Eagle}, two probe-based schedulers which use probe sampling. We evaluate the sensitivity of Peacock to probe rotation and probe reordering. We use both simulation for large clusters of 10k, 15k, and 20k workers and real implementation for 100 workers.\\
\textbf{Environment.} We implemented an event-driven simulator and also all three algorithms within it to fairly compare them for large scale cluster sizes. In addition, we implemented Peacock as an independent component using Java and also a plug-in for Spark~\cite{USENIX:Zaharia} written in Scala. We used Sparrow and Eagle source codes for the distributed experiments.\\
\textbf{Workload.} We utilize traces of Google~\cite{GoogleTraceWebsite,ACM:Charles}. Invalid jobs are removed from the Google traces and Table~\ref{table:datasetprops} gives the specification of the pruned traces. To generate  \textit{average} cluster workloads, job arrival time follows a Poisson process with a mean job inter-arrival time that is calculated based on expected average workload percentage, mean jobs execution time, and mean number of 
\begin{table}[htp]
\setlength{\tabcolsep}{3pt}
\caption{Workloads general properties}
\label{table:datasetprops}
\centering 
\begin{tabular}{  c | c | c | c  } 
\toprule
Workloads &  Jobs Count & Tasks Count & Avg Task Duration \\
\toprule
  Google & 504882 & 17800843 & 68  \\	
\toprule
\end{tabular}
\end{table}
tasks per job. Since jobs are heterogeneous, the workload and expected average percentage vary over time. We consider 20\%, 50\%, and 80\% as light  and 100\%, 200\%, and 300\% as heavy cluster workloads.     \\
\textbf{Parameters.} The estimated task runtime  is computed as the average of job task durations. Each worker runs one task at a time, which is analogous to having multi-slot workers, each is served by a separate queue. The results are the average of a number of runs. Error bars are ignored due to stable results of different runs. We set rotation interval to 1s and network delay to 5ms for simulation experiments. Eagle relies on several static parameters. For fair comparison, we use the values used in the paper~\cite{Eagle} even though any algorithm relying on static values may not be appropriate under dynamic workloads.\\
\textbf{Performance Metrics.} We measure the average job completion times, cumulative distribution function of job completion times, and the fraction of jobs that each algorithm completes in less time comparatively, to appraise how efficiently Peacock mitigates the \emph{Head-of-Line blocking}.
\begin{algorithm}[ht]
\caption{Enqueue Probe submitted by scheduler or rotated by predecessor}
\label{algo:enqueue}
\begin{algorithmic}[1]
\Procedure{enqueueProbe(\textit{p})}{}
    \State{\textit{$\gamma_{p}$ $\leftarrow$ $\delta$ + $\alpha$}}
	\For{\texttt{q in reversed waitingProbes}}
		\If{$\lambda_{p}$ $\geq$ $\lambda_{q}$}
           \If{$\theta_{p}$ $\leq$ $\theta_{q}$ AND $\lambda_{q}$ + $\mu_{q}$ + $\theta_{p}$ $\leq$ $\tau$}
                \State{$\gamma_{p}$ = $\gamma_{p}$ - $\theta_{q}$}
             \Else
                 \State{\textsc{placeOrRotate}(p) ; decided = true ;  break;}
           \EndIf
             \Else 
           \If{$\theta_{q}$ $\leq$ $\theta_{p}$ AND $\tau$ + $\gamma_{p}$ $\leq$ $\lambda_{p}$ + $\mu_{p}$}
                \State{\textsc{placeOrRotate}(p) ; decided = true ; break;}
             \Else
                \State{$\gamma_{p}$ = $\gamma_{p}$ - $\theta_{q}$}
           \EndIf
		\EndIf
	 \EndFor
     \If{Not decided}
         \State{waitingProbes.add(P , 0) ; $\alpha$ = $\alpha$ + $\theta_{p}$}
     \EndIf
  \While{waitingProbes.size() $\geq$ $\phi$ OR $\alpha$ $\geq$ $\omega$}
    \State{q = waitingProbes.removeLast() ;$\alpha$ = $\alpha$ - $\theta_{q}$ ; rotatingProbes.add(q)}
  \EndWhile
\EndProcedure
\Procedure{placeOrRotate(\textit{p})}{}
     \If{$\tau$ + $\gamma_{p}$  $\leq$ $\lambda_{p}$ + $\mu_{p}$ OR $\lambda_{p}$ + $\mu_{p}$ $\leq$ $\tau$}
         \State{waitingProbes.add(P) ; $\alpha$ = $\alpha$ + $\theta_{p}$}
       \Else
          \State{rotatingProbes.add(p)}
     \EndIf
\EndProcedure
\end{algorithmic}
\end{algorithm}
\section{Experimental Results}\label{sec:Simulation}
We deploy our algorithm within an event-driven simulator and a real distributed experiment  to evaluate Peacock in different loads and cluster sizes.
\subsubsection{Comparing Peacock Against Sparrow.}
\begin{figure*}[htp!]
  \centering
 \subfigure[Google-Heavy]{\includegraphics[scale=0.4]{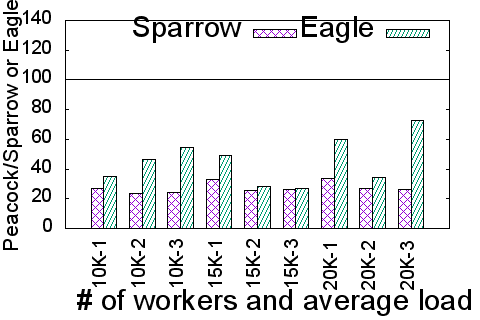}}\quad
 \subfigure[Google-Light]{\includegraphics[scale=0.4]{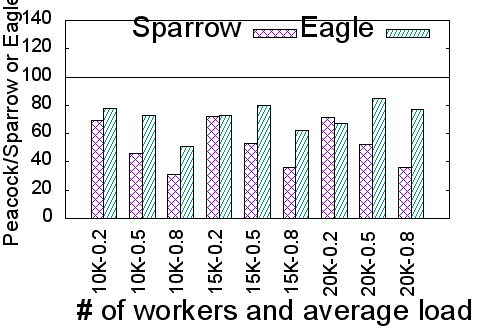}}\quad
\caption{Average Job Completion times for heavy and light load scenarios.}
\label{fig:ajct}
\end{figure*}

\begin{figure*}[htp!]
  \centering
   \subfigure[Google-300\%]{\includegraphics[scale=0.4]{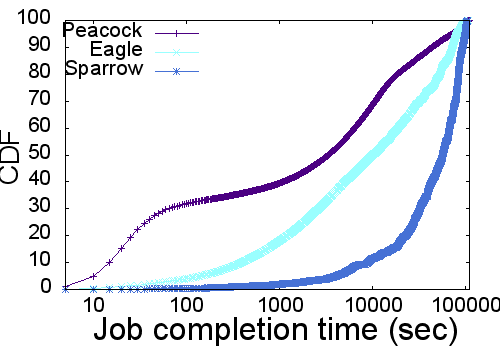}}\quad
     \subfigure[Google-50\%]{\includegraphics[scale=0.4]{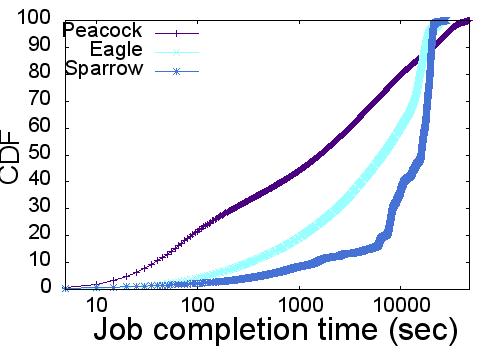}}\quad
   \caption{Cumulative distribution function of Jobs completion times.10000 workers.}
   \label{fig:CDF}
\end{figure*}

\begin{figure*}[htp!]
  \centering
  \subfigure[Google-Light]{\includegraphics[scale=0.3]{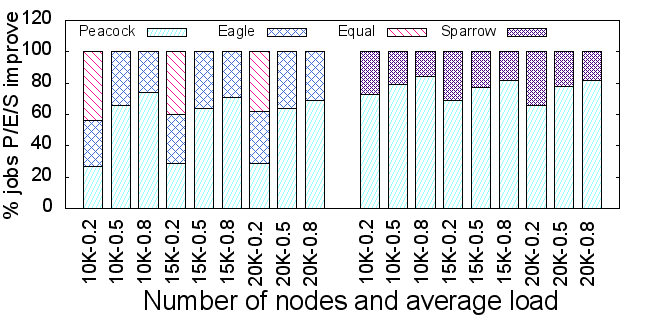}}\quad
  \subfigure[Google-Heavy]{\includegraphics[scale=0.3]{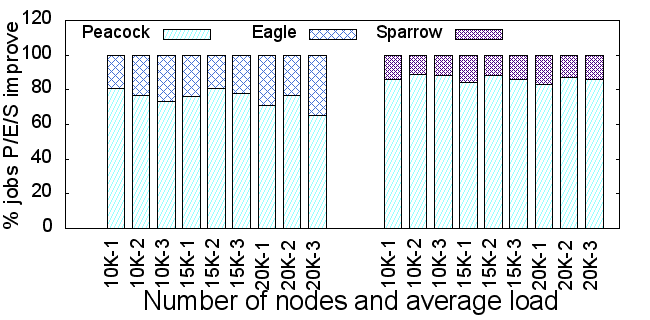}}\quad
  \caption{Fraction of jobs with shorter completion time.}
  \label{fig:percentage}
\end{figure*}

Figure~\ref{fig:ajct} shows that Peacock achieves  better average jobs completion times than Sparrow under all loads and with all cluster sizes. Peacock outperforms the alternatives under heavy loads. The reason is that \emph{Head-of-Line blocking} is reduced (i)~locally in each worker by our reordering and (ii)~collaboratively between workers by balancing the distribution of probes through both probe rotation and reordering. In light loads, the improvement is mostly due to probe rotation and rarely due to the reordering. Furthermore, Sparrow only uses batch sampling that does not handle workload heterogeneity. Figure~\ref{fig:CDF} shows that Peacock, unlike Sparrow, is job-aware in the sense that it reduces the variance of task completion times for each job. Beside probes rotation and reordering, the way that Peacock assigns threshold value for jobs appears effective. Figure~\ref{fig:percentage} shows that Peacock significantly outperforms Sparrow when comparing jobs individually. Under a 20\% load, Sparrow shows better percentage than other loads because two samplings in Sparrow get empty slots faster than one sampling of Peacock even though probe rotation helps Peacock outperform Sparrow under other loads. We now provide some more detailed information. Figure~\ref{fig:ajct} shows Peacock executes jobs in average between 13\% to 77\% faster than Sparrow in all settings. Figure~\ref{fig:CDF}(b) shows in 50\% load, Sparrow only completes 2.2\% jobs in less than 100 seconds while Peacock completes 21.6\% jobs at the same time. In Figure~\ref{fig:CDF}(a) and under the 300\% load, Sparrow executes 0.3\% jobs less than 100 seconds while it is 31.8\% for Peacock. Figure~\ref{fig:percentage} shows that Peacock executes between 66\% to 91\% of jobs faster than Sparrow. 
\subsubsection{Comparing Peacock Against Eagle.}
Eagle is a hybrid probe-based sampling scheduler which divides jobs statically into two sets of long and short jobs. A centralized node schedules long jobs and a set of independent schedulers using batch sampling to schedule short jobs. The cluster is divided into two partitions, one is dedicated to short jobs and the other is shared for all jobs. Eagle mitigates~\emph{Head-of-Line blocking} using re-sampling technique and a static threshold-based queue reordering. Figure~\ref{fig:ajct} shows that Peacock outperforms Eagle in average jobs completion times in all loads. It is because the continuous and deterministic probe rotations through elastic queues along with the workload-aware probe reordering in Peacock outperforms a randomized re-sampling along with a static probe reordering through unbounded queues in Eagle. In Figure~\ref{fig:CDF}, we see that Peacock executes jobs in lower latency than Eagle. Figures~\ref{fig:ajct} shows, Peacock completes execution of jobs in average 16\% to 73\% faster than Eagle. Figure~\ref{fig:percentage} shows Peacock executes between 54\% to 82\% of jobs faster than Eagle. Interestingly, we see that under 20\% load, the percentage of jobs have identical completion time in both Eagle and Peacock. Figure~\ref{fig:CDF} shows that Peacock executes however a high percentage of jobs with lower latency than Eagle.
\begin{wrapfigure}{l}{6.7cm}
  \includegraphics[scale=0.4]{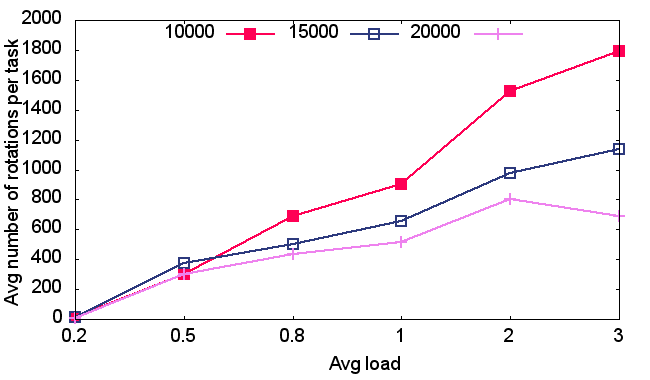}
  \caption{Avg number of rotations per probe}
  \label{fig:numberofrotations}	
\end{wrapfigure} 
\subsubsection{How much is the number of probe rotations per task influenced by cluster sizes and loads?}\label{sec:periods}
We investigate the average number of probe rotations
per task for Google trace.
We observe by increasing the cluster size that the number of rotations decreases. For example, for 80\% load, the number of rotations for 10K, 15K, and 20K nodes are 901, 656, and 513 respectively.  Also, for higher loads, at 300\%, the number of rotations are 1791, 1140, and 692 for 10K, 15K, and 20K, respectively. The larger the cluster size, the lower the number of redundant rotations. It indicates that probe rotation does not hurt the scalability and hence Peacock can be deployed on large scale clusters. In addition, by increasing the load, there is a reduction in the number of rotations for all 3 cluster sizes. The heavier loads trigger a higher number of rotations than lighter loads. For 10K the number of rotations are 17, 299, 689, 901, 1523, and 1791 for 20\%, 50\%, 80\%, 100\%, 200\% and 300\% loads respectively.   
\subsubsection{Sensitivity to Probe Rotation.}
\begin{figure*}[htp!]
  \centering
  \subfigure[w/o rotating AJCT]{\includegraphics[scale=0.4]{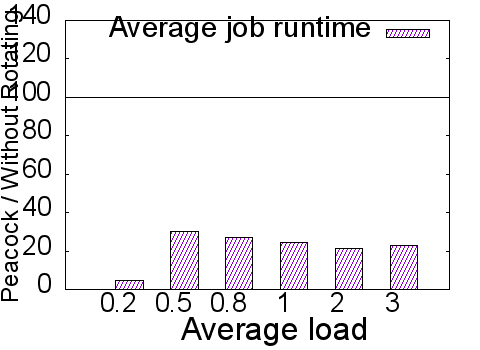}}\quad
  \subfigure[w/o rotating Percentiles]{\includegraphics[scale=0.4]{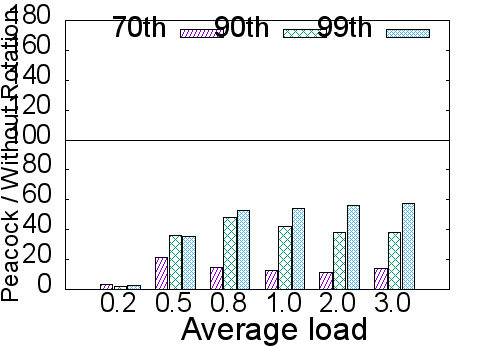}}\quad
  \subfigure[w/o reordering AJCT]{\includegraphics[scale=0.4]{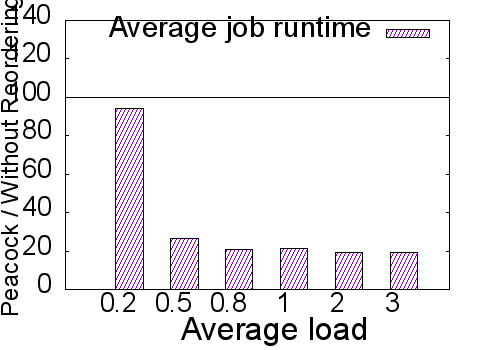}}\quad
  \subfigure[w/o reordering Percentiles]{\includegraphics[scale=0.4]{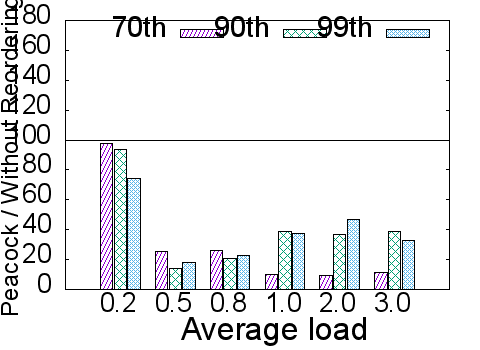}}\quad
  \caption{Peacock versus  w/o  probes rotation or probes reordering. Google trace.}
  \label{fig:RotatingReordering}
\end{figure*}

We analyze the effectiveness of probe rotation on the performance of Peacock. Figures~\ref{fig:RotatingReordering}(a) and~\ref{fig:RotatingReordering}(b) reveal that the performance of Peacock stems from probe rotation technique on all loads. From Figure~\ref{fig:RotatingReordering}(a), we see the average job completion time is negatively increased between 70\% to 95\% in all loads in comparison with complete Peacock version because probe rotation mitigates \emph{Head-of-Line blocking}. Specifically, in light loads, probe rotation balances load between workers which result in increasing the cluster utilization and greatly reducing the formation of long-length queues. In heavy loads, due to the existence of long-length queues, Besides balancing the load between workers through probe reordering, Peacock uses probe rotation to mitigate \emph{Head-of-Line blocking}. Figure~\ref{fig:RotatingReordering}(b) shows that 70\% and 90\% percentiles in the high loads perform better than the same percentiles for the light loads. It indicates that under high load probe reordering and probe rotation collaboratively mitigates \emph{Head-of-Line blocking} while under light load the performance of probes rotation is crucial as there is no long queues to apply probes reordering.
\subsubsection{\textbf{Sensitivity to Probe Reordering.}}
Probe reordering is more influential when the cluster is under a high load since workers have long queues when they are at high load. Thanks to the novel starvation-free reordering algorithm in which it allows jobs to bypass longer jobs. The result in Figure~\ref{fig:RotatingReordering}(c) approves this fact wherein average job completion time for Peacock without reordering component is close to the original Peacock for 20\% load while by increasing load, we observe an increasing difference in average job completion time (the biggest difference is 81\% for loads 200\% and 300\%). From  Figure~\ref{fig:RotatingReordering}(d) we can conclude that reordering causes most of jobs to be executed faster. It shows an improvement of 90\% in 70\% percentile for loads 100\%, 200\%, and 300\% while load 50\% with 76\% and 74\% improvements has the best percentiles in 90\% and 99\%. As expected there is no significant difference for load 20\% as there is no waiting probes in queues most of the time. It is obvious that the elimination of this component significantly increases the chance of having \emph{Head-of-Line blocking}. 
\subsubsection{Implementation Results.} \label{sec:realImplementation}
\begin{figure*}[htp!]
  \centering
  \subfigure[Avg Job completion time]{\includegraphics[scale=0.3]{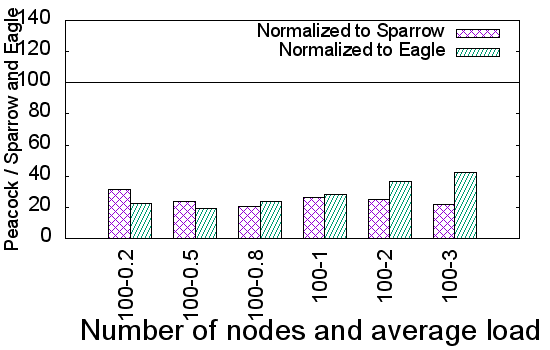}}\quad
  \subfigure[\% of shorter completed jobs]{\includegraphics[scale=0.3]{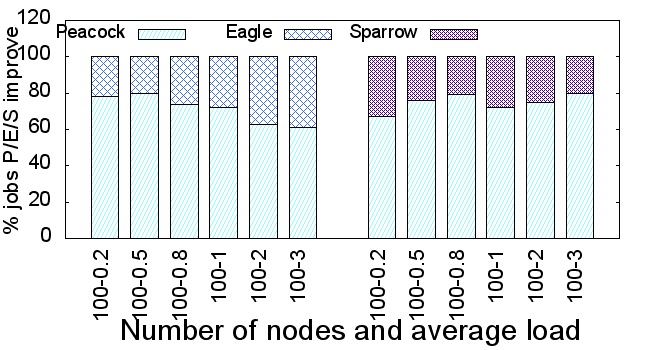}}\quad
  \caption{Distributed experiments for heavy and light workloads.}
  \label{fig:realImplementation}
  \end{figure*}

We implement Peacock as an independent component using Java and  a plug-in for Spark~\cite{USENIX:Zaharia} written in Scala. We run experiments on 110 nodes consisting of 100 workers and 10 schedulers. To keep it traceable, we sample 3200 jobs of Google trace and we convert task durations from seconds to milliseconds. We implement a Spark job called \emph{sleep task}. The current thread sleeps for a duration equals to task duration to simulate the execution time that each task needs. The method for varying the load is the same as the simulation experiments described in section~\ref{sec:Evaluation}. We run real implementations of Sparrow and Eagle with the same specifications to compare Peacock against them. Figure~\ref{fig:realImplementation}(a) presents average job completion time at both light and heavy
loads. The result shows that Peacock significantly outperforms both the algorithms in all loads. Peacock outperforms Sparrow with an at most 80\% improvement under the 80\% load and at least a 69\% improvement under the 20\% load scenario. Moreover, compared to Eagle, the maximum improvement reaches 81\% when the load is 50\% and the least enhancement is 57\% for the load 300\%. Figure~\ref{fig:realImplementation}(b) shows the fraction of jobs that each algorithm runs in less time. Again we can see that Peacock runs higher percentage of jobs faster than both Sparrow and Eagle.  
\section{Related work}\label{sec:Relatedwork} 
Original schedulers are usually designed in a centralized manner~\cite{Reservation-based,Delayscheduling,Jockey,Bistro,Quincy,Borg} and are a computationally expensive optimization problem. Such algorithms may increase scheduling times and lead to scalability problems. Distributed and hybrid schedulers are proposed to resolve the problem. Sparrow~\cite{ACM:Ousterhout} is a distributed scheduler using batch sampling and late binding techniques to be scalable and offer low latency. However, it faces challenges in highly loaded clusters due to the lack of \emph{Head-of-Line blocking} mitigation. Hawk~\cite{Hawk} and Eagle~\cite{Eagle} are hybrid schedulers that augment Sparrow to mitigate \emph{Head-of-Line blocking}. A centralized scheduler schedules long jobs and distributed schedulers handles short jobs. Both divide jobs statically into long and short categories, splits workers into two partitions statically, and allocate one partition to short jobs and another to both types of jobs. To mitigate the \emph{Head-of-Line blocking} in Hawk, idle workers steal short tasks that get stuck behind long jobs. Instead, Eagle shares information among workers called Succinct State Sharing, in which the distributed schedulers are informed of the locations where long jobs are executing. Eagle also proposes a Shortest Remaining Processing Time reordering technique to prevent starvation. Unfortunately, Eagle relies strongly on static parameters which limits its practicality and does not perform well under light loads. In Mercury~\cite{Mercury}, jobs are divided into two sets, either served centrally with best effort or scheduled by distributed schedulers. It uses a load shedding technique to re-balance load on workers. Mercury does no cope with the \emph{Head-of-Line blocking} and faces scalability issues when there are a large number of guaranteed jobs waiting to be scheduled. Apollo~\cite{Apollo} relies on shared states. Jobs are homogeneous and scheduled with the same policy. A centralized manager maintains a shared state updated by connecting with nodes. Unlike Apollo, Peacock imposes a tiny global information not relying on central coordination.
\section{Conclusion}\label{sec:Conclusion}
We presented Peacock, a new distributed probe-based scheduler for large scale clusters. Peacock mitigates the \emph{Head-of-Line 
blocking} by combining probe rotation 
through the elastic queues with a novel probe reordering. Peacock organizes workers into a ring overlay network and regulates probes to move between workers through the elastic queues of workers to handle workload fluctuations. We showed that Peacock outperforms state-of-the-art probe-based schedulers in various workloads through simulation and realistic distributed experiments.  
%

%
%


\begin{thebibliography}{1}
\bibitem{USENIX:Zaharia}
Zaharia, Matei, et al. "Resilient distributed datasets: A fault-tolerant abstraction for in-memory cluster computing." In: NSDI (2012)
\bibitem{GoogleTraceWebsite}
 GoogleTraceWebsite. Google cluster data. https://code.google.com/p/googleclusterdata/.
\bibitem{ACM:Ousterhout}
Ousterhout, Kay, et al. "Sparrow: distributed, low latency scheduling." In: SOSP. (2013)
\bibitem{Eagle}
Delgado, P., Didona, D., Dinu, F., \& Zwaenepoel, W. (2016, October). Job-aware scheduling in eagle: Divide and stick to your probes. In: SOCC. (2016)
\bibitem{Hawk}
Delgado, Pamela, et al. "Hawk: Hybrid Datacenter Scheduling." USENIX Annual Technical Conference. 2015.
\bibitem{ACM:Tumanov}
Tumanov, Alexey, et al. "TetriSched: global rescheduling with adaptive plan-ahead in dynamic heterogeneous clusters." In: EuroSys. (2016)
\bibitem{ACM:Rasley}
Rasley, Jeff, et al. "Efficient queue management for cluster scheduling." Proceedings of the Eleventh European Conference on Computer Systems. ACM, 2016.
\bibitem{Mercury}
Karanasos, Konstantinos, et al. "Mercury: Hybrid Centralized and Distributed Scheduling in Large Shared Clusters." USENIX Annual Technical Conference. 2015.
\bibitem{Quincy}
Isard, Michael, et al. "Quincy: fair scheduling for distributed computing clusters." In: SOSP. (2009)
\bibitem{Mesos}
Hindman, Benjamin, et al. "Mesos: A Platform for Fine-Grained Resource Sharing in the Data Center." NSDI. Vol. 11. No. 2011. 2011.
\bibitem{Apollo}
Boutin, Eric, et al. "Apollo: Scalable and Coordinated Scheduling for Cloud-Scale Computing." OSDI. Vol. 14. 2014.
\bibitem{Jockey}
Ferguson, Andrew D., et al. "Jockey: guaranteed job latency in data parallel clusters." In: EuroSys (2012).
\bibitem{Delayscheduling}
Zaharia, Matei, et al. "Delay scheduling: a simple technique for achieving locality and fairness in cluster scheduling." In: EuroSys (2010).
\bibitem{Reservation-based}
Curino, Carlo, et al. "Reservation-based scheduling: If you're late don't blame us!." Proceedings of the ACM Symposium on Cloud Computing. ACM, 2014.
\bibitem{Bistro}
Goder, Andrey, Alexey Spiridonov, and Yin Wang. "Bistro: Scheduling Data-Parallel Jobs Against Live Production Systems." In: USENIX ATC. (2015)
\bibitem{Borg}
Verma, Abhishek, et al. "Large-scale cluster management at Google with Borg." Proceedings of the Tenth European Conference on Computer Systems. ACM, 2015.
\bibitem{ACM:Charles}
Reiss, Charles, et al. "Heterogeneity and dynamicity of clouds at scale: Google trace analysis." In: SOCC. (2012)
\bibitem{SCOPE}
Zhou, Jingren, et al. "SCOPE: parallel databases meet MapReduce." The International Journal on Very Large Data Bases 21.5 (2012): 611-636.
\bibitem{chord}
Stoica, Ion, et al. "Chord: A scalable peer-to-peer lookup service for internet applications." ACM SIGCOMM Computer Communication Review 31.4 (2001): 149-160.
\bibitem{YahooTrace}
Chen, Yanpei, et al. "The case for evaluating MapReduce performance using workload suites." In: MASCOTS. (2011) 
\end{thebibliography}
\end{document}